DEC-TR-342
# On Caching Out-of-Order Packets in Window Flow Controlled Network

Raj Jain

Eastern Research Laboratory
Digital Equipment Corporation
77 Reed Road
Hudson, Massachusetts 01749

## Abstract
In window flow controlled networks, if a packet is lost the destination has to decide whether to save (cache) subsequent out-of-order packets. Also, the source has to decide whether to send just one packet or to send all packets following it. This leads to four different types of caching schemes. Simulations show, against our immediate intuition, that regardless of whether the destination is caching or not, the source should retransmit only one packet. This paper describes the alternatives to, and provides justification for, schemes used in Digital Network Architecture and ARPAnet TCP.

# 1 Introduction

The problems of timeout, caching out-of-order packets, and congestion are related. Congestion is the prime cause of packet losses, which cause timeouts and necessitate caching out-of-order packets. Not caching out-of-order packets intensifies the impact of congestion and packet loss. Finally, a bad timeout algorithm (too quick timeouts) may also lead to congestion.

While investigating the problem of congestion in Digital Network Architecture (DNA), we discovered several interesting phenomena about these three problems. DNA (more commonly known by its product name DECnet) is a typical packet switching network architecture very similar to ARPAnet [2], and closely follows ISO's OSI reference model [4]. We hope that our experiences in DNA will help other networking architects.

The organization of the paper is as follows. We first list the key design decisions related to caching out-of-order packets and describe the schemes used in DNA and ARPAnet. We then describe other alternatives that were proposed with the hope that they would improve performance. The simulation shows that the current schemes are preferable. Finally, the justification for this conclusion is presented.

# 2 Design Decisions Related to Caching

When designing an end-to-end protocol, an architect has to make three decisions regarding to caching out-of-order packets.

1. Whether the destinations should cache out-of-order packets.

2. On a timeout, whether the source should retransmit all packets following the last acknowledged packet.

3. Whether the connection set-up protocol should include information about the caching; i.e., whether each side should know if the other is caching out-of-order packets.

In DNA, as originally designed [1], the decisions were:

1. The caching is optional.

2. The source should send just one packet, the one that timed out.

3. At connection setup, the nodes do not exchange information about caching behavior.

It appears that caching would save many retransmissions. However, it also results in many buffers being tied up at the destination for the duration of the timeout. These buffers could be used by other connections on the node. Therefore, most implementations of DNA do not provide caching. These decisions are typical of those in other networks as well. ARPAnet TCP protocol, for example, has exactly the same decisions: caching is optional sources retransmit only one segment [2], and the caching behavior of the destination is not known to the source.

Not providing caching leads to an infinite cycle of retransmissions, if the flow control window is more than one. In this case, as shown in Figure 1, loss of a single packet can start a loop in which each packet is transmitted twice, the timer algorithm diverges, and throughput quickly drops to zero [3]. This happens because all subsequent packets are considered out-of-order.

This led us to see if it would help to tell the source whether the destination is caching. Our simulation shows (against our immediate intuition) that the present scheme of not telling the source about the real status of caching is better. The explanation follows.

## 3 Four Types of Caching Schemes

The two choices for a source are to be optimistic or to be pessimistic. Being pessimistic means assuming that the destination is not caching the out-of-order packets. Being optimistic means that the destination is caching out-of-order packets. Similarly, there are two types of destinations: some that cache and some that do not cache out-of-order packets. This leads to four different types of out-of-order caching (OOC) schemes, as shown in Figure 2.

## 3.1 OOC-1 (Source Optimistic, Destination Not Caching)

In this case, if a packet is lost, the source retransmits just that packet and hopes that all subsequent packets have been cached by the destination. As shown under OOC-1 in Figure 2a, packet 1 is sent but lost. The source continues to send packets 2, 3, and 4 before finding out that the acknowledgment for packet 1 has not arrived, i.e., packet 1 has timed out. The source it retransmits packet 1. Then if the flow control window permits, the scource continues to send packets 5, 6, etc. At the destination, packet 1 does not arrive, so packets 2, 3, and 4 are dropped because they are out-of-order. When packet 1 arrives, it is acknowledged. A permission to send an additional packet may be issued to the source, which proceeds to send packet 7. Later, the source times out on packet 2, retransmits it, then times out on packet 3, and so on.

It is obvious that after a single packet loss, all subsequent packets will have to be transmitted twice. What is not so obvious is the fact that each timeout also results in an increased round trip delay estimate. This increases the next timeout interval, which increases the next estimate, and so on. The timer algorithm diverges, and throughput drops to zero- just because one packet was lost.

## 3.2 OOC-2 (Source Pessimistic, Destination Not Caching)

It may appear that one way to fix the above problem is for the source to retransmit all outstanding packets on a timeout. This is what OOC-2 does. In the example OOC-1, when packet 1 times out, the source retransmits packets 1 through 4.

## 3.3 OOC-3 (Source Pessimistic, Destination Caching)

This scheme is doubly cautious. Not only does the destination cache all out-of-order arrivals, but on a timeout, the source also retransmits all outstanding packets. These are later dropped at the destination as being duplicates. The destination sends an acknowledgment for packet 4 as soon as it receives the second copy of packet 1.

## 3.4 OOC-4 (Source Optimistic, Destination Caching)

This is obviously the best case. The source sends just the packet that times out, and the destination caches packets previously received and acknowledges all of them as soon as it receives the lost packet.

# 4 Why are the Optimistic Schemes Are Better

Telling the source about the caching status of the destination is equivalent to choosing OOC-2 and OOC-4. Not telling is equivalent to choosing one of the two subsets {OOC-1, OOC-4} or {OOC-2, OOC-3}. Our simulation results recommend choosing {OOC-1, OOC-4}, the optimistic subset.

Before simulating these possibilities, we reasoned that the performance order of these four alternatives should be:

$$OOC - 1 < OOC - 2 < OOC - 3 < OOC - 4$$

Our reasoning was as follows. OOC-2 should be better than OOC-1 because there are fewer timeouts. OOC-3 should be better than OOC-2 because other things being equal, a caching destination is better than a non-caching destination. OOC-4 should be better than OOC-3 because there are no duplicates.

The simulation confirmed our intuition, although only partly. All the above arguments are valid for lightly loaded networks. However, for congested networks during which the packets are actually lost and the caching plays an important role, the simulation results showed that optimistic schemes perform better than pessimistic schemes even if the destination is not caching. During congestion

$$OOC - 2 << OOC - 1 \text{ and}$$

$$OOC - 3 << OOC - 4$$

Although the original source of this discovery was a simulation model, we do need not to go into the details of the simulation here. The following argument explains the result.

In a congested network, every extra packet in the network only makes the congestion worse. If a source had a window size of $C$ packets, on a single packet loss, a pessimistic scheme would inject $C$ packets in to the network. All of these would probably also be lost because there may not be enough buffers available at intermediate nodes. Also, if there were $n$ successive alarms or timeouts (some of which might be false alarms), a pessimistic scheme would inject a total of $(1+n)C$ packets into the network, as opposed to only $C+n$ packets for optimistic schemes. Thus pessimistic schemes are more likely to aggravate congestion.

Since the main reason for having a caching scheme is to help improve the performance during congestion, the conclusion is to use optimistic schemes.

## 5 Conclusion

If the major cause of packet loss is congestion, one should be conservative in injecting more packets into the network. A loss, assuming that the network parameters are properly tuned, indicates onset of the congestion. Therefore, the best strategy is to retransmit just one packet that timed out, regardless of whether the destination is caching out-of-order packets or not. Also, since the source behavior is the same regardless of the destination's decision to cache, it does not pay off to pass on the destination's decision to the source at the connection set up time.

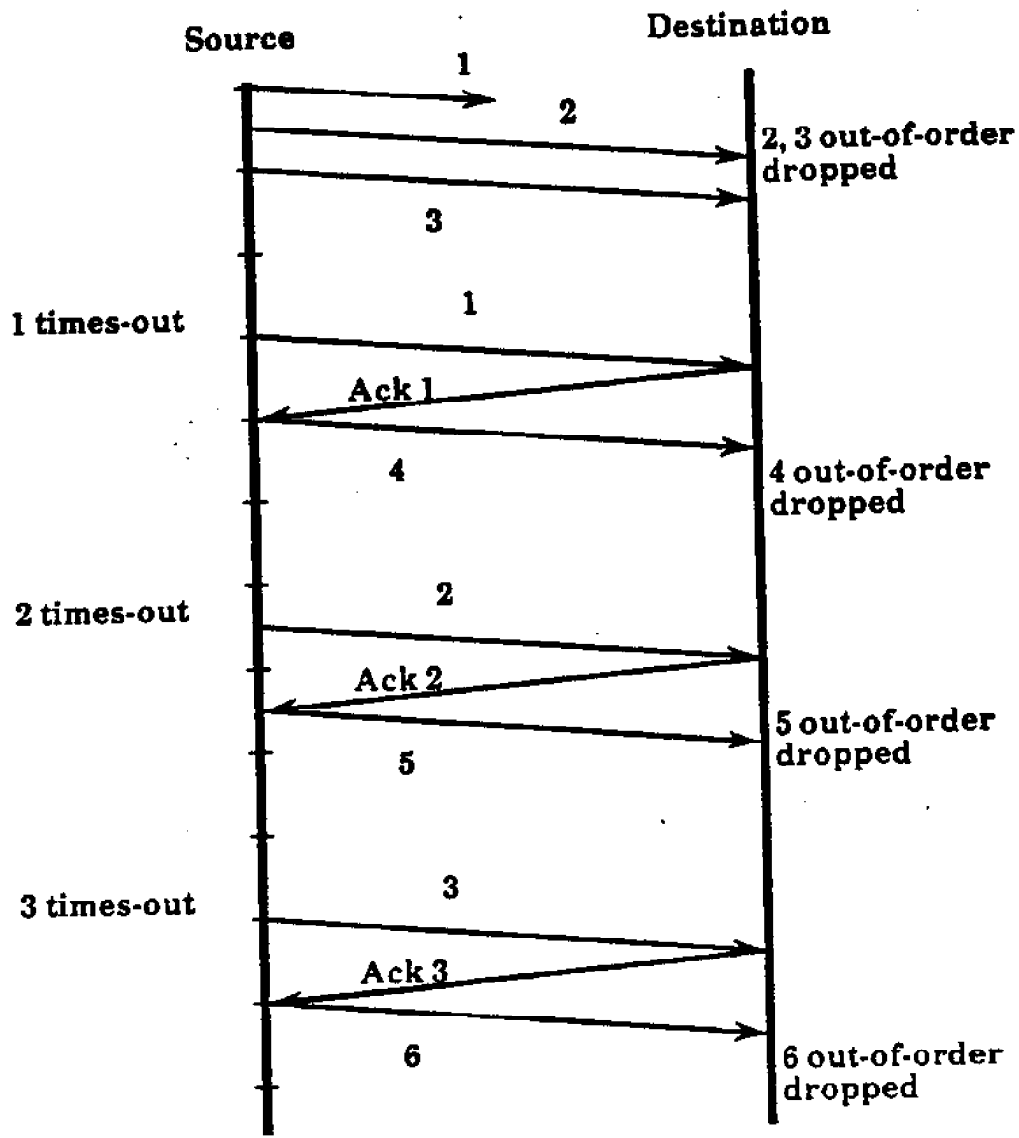

Figure 1: With a flow control window greater than one, the loss of a single packet may lead to all subsequent packets being transmitted twice if the destination does not cache out-of-order packets. The timer algorithm diverges and the throughput drops to zero.

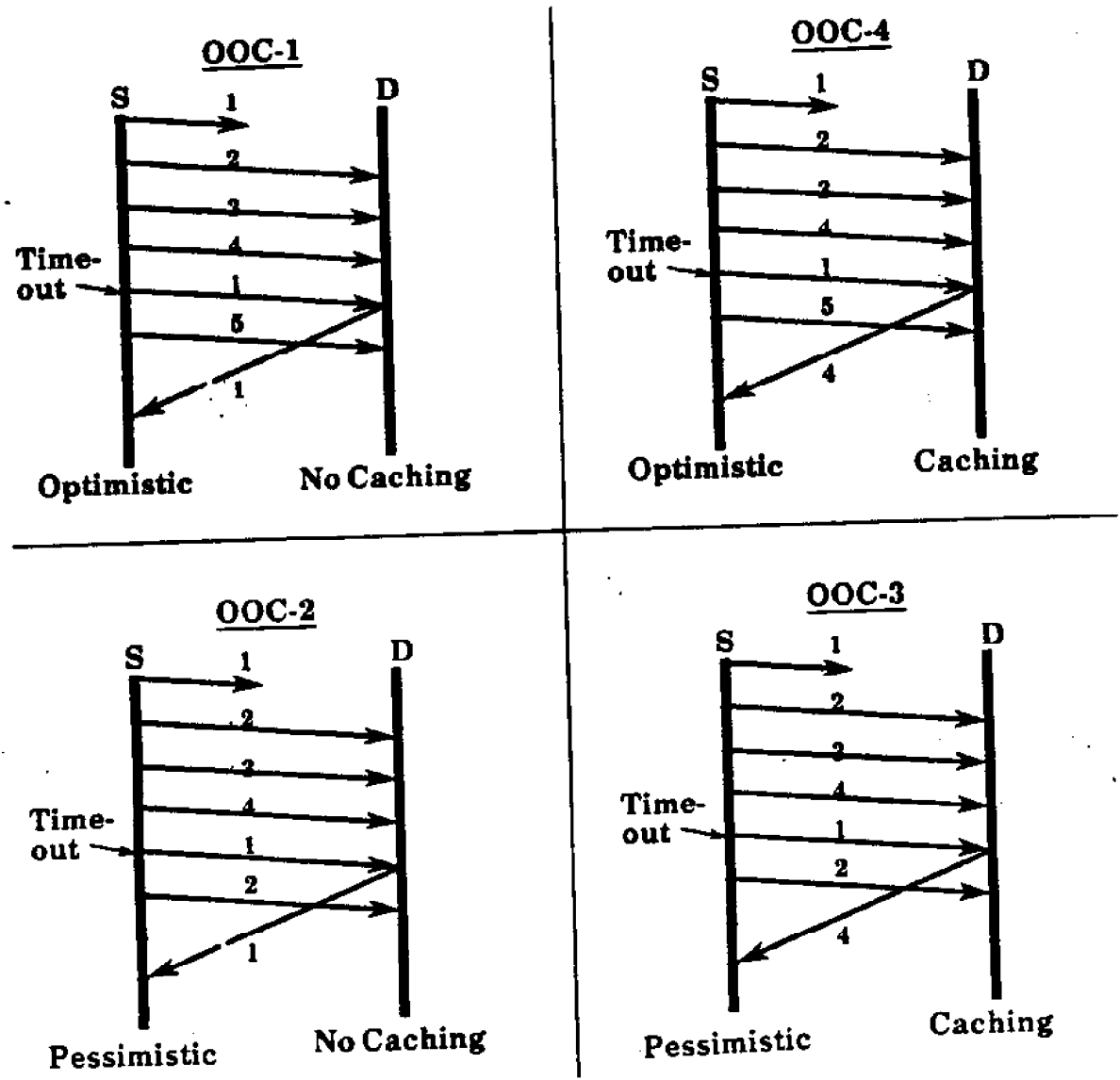

Figure 2: There are four schemes for caching out-of-order packets. The destination may or may not cache, and the source may or may not assume that the destination is caching. Simulation shows that during congestion, it is better if the source is optimistic and does not inject too many packets into the network.